\documentclass[twocolumn,floatfix,prl,showpacs]{revtex4}%

\usepackage{epsfig}
\usepackage{dcolumn}
\usepackage{bm}
\usepackage{amsmath}
\usepackage{graphicx}

\begin{document}

\title{Narrow band microwave radiation from a biased single-Cooper-pair transistor}

\author{O. Naaman}

\author{J. Aumentado}
\email{jose.aumentado@boulder.nist.gov}
\affiliation{National Institute of Standards and Technology, 325 Broadway, Boulder, Colorado 80305, USA}

\date{September 19, 2006}

\begin{abstract}
We show that a single-Cooper-pair transistor (SCPT) electrometer emits narrow-band microwave radiation when biased in its sub-gap region. Photo activation of quasiparticle tunneling in a nearby SCPT is used to spectroscopically detect this radiation, in a configuration that closely mimics a qubit\textendash electrometer integrated circuit. We identify emission lines due to Josephson radiation and radiative transport processes in the electrometer, and argue that a dissipative superconducting electrometer can severely disrupt the system it attempts to measure.
\end{abstract}

\pacs{85.25.Cp, 74.50.+r, 03.67.Lx, 85.60.Gz}
\maketitle

The implementation of a quantum computer in the solid state requires, aside from the quantum bit (qubit) itself, an integrated readout device. It must be fast and sensitive, and present the qubit with a minimal source of decoherence. In the Cooper-pair box qubit \cite{Nakamura99,Vion02}, a natural choice for a readout device is the rf single-Cooper-pair transistor (SCPT) electrometer\textemdash an rf single-electron transistor (SET) operating in the superconducting state \cite{Devoret00,Aassime01,Lehnert03}. However, there has been growing concern recently that the voltage-biased, and thus \textit{dissipative}, operation of the superconducting rf-SET electrometer makes it less than ideal in measuring quantum circuits \cite{Turek05,Devoret04,Mannik04}. Not only does the electrometer provide a dissipative environment that may relax and dephase the qubit, but the nontrivial backaction noise associated with the various transport mechanisms through the electrometer may also excite the qubit and even lead to a population inversion \cite{Clerk02,Clerk05}. Despite the availability of new dispersive readout schemes \cite{Sillanpaa04,Wallraff05,Siddiqi06}, the dissipative superconducting rf-SET is still widely used as an electrometer in quantum circuits \cite{Lu03,Buehler06,Delsing06}. There is a strong need, therefore, to experimentally determine how the dissipative superconducting electrometer affects the system it measures, and specifically, what are the components of the electrometer's emission spectrum.

In this Letter, we report on our measurements of the emission spectrum of a biased SCPT. While narrow-band microwave radiation from a biased SCPT has been previously observed using photon-assisted tunneling in a strongly coupled SIS junction detector \cite{Deblock03}, those measurements were done in a limited range of the electrometer's operating point near the Josephson-quasiparticle peak. We spectroscopically measured the radiation emitted from the electrometer as a function of its operating point over a wide range of voltage and charge bias conditions, throughout its sub-gap region. We detected this radiation by photon-assisted quasiparticle (QP) tunneling in a nearby SCPT. 

\begin{figure}[t]
\epsfxsize=3.2in
\epsfbox{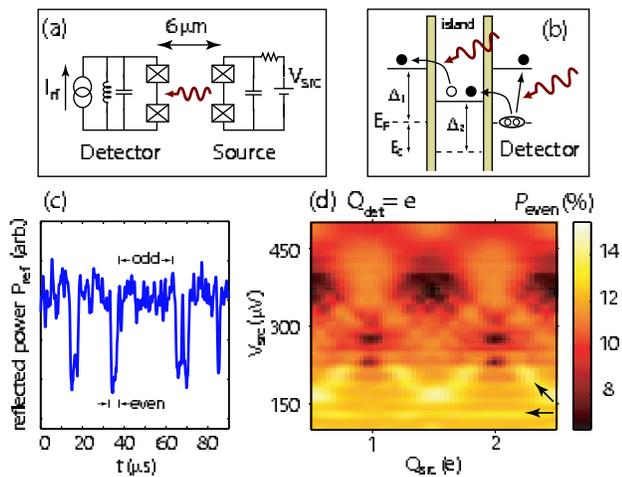}
\caption{\label{circuit} (color online). (a) A schematic of the experimental setup. The detector SCPT is embedded in a resonator and measured by rf reflectometry. (b) A cartoon of photon-assisted QP tunneling in the detector, $Q_\textrm{det}=e$. (c) Reflected rf power P$_\textrm{ref}$ vs.~time at $Q_\textrm{det}=e$. $f=510$ MHz, P$_\textrm{in}=3.2$ fW ($-115$ dBm). (d) Detector even-state probability vs.~source operating point at $Q_\textrm{det}=e$. Arrows indicate ridges with enhanced even-state probability. No change in $P_\textrm{even}$ was observed for $|V_\textrm{src}|<100\:\mu$V.}
\end{figure}

Our experimental setup is shown in Fig.~\ref{circuit}(a). The two SCPTs were co-fabricated by double-angle Al deposition, and were separated by $\sim6\;\mu$m. As in Ref.~\cite{Mannik04}, the islands of the two devices are not coupled directly; their leads, however, provide stray coupling at microwave frequencies \cite{NoteCoupling}. In both devices the charging energy is $E_C=e^2/2C\sim 170\;\mu e$V ($C$ is the total island capacitance) and the normal state resistance is $R_N$$\sim$22\;k$\Omega$. We oxygen-doped one of the Al layers to increase its superconducting gap \cite{Aumentado04}, $\Delta_1=225\;\mu$eV (20 nm thick), the other Al film was 40 nm thick with $\Delta_2=190\;\mu$eV. In the device on the right of Fig.~\ref{circuit}(a), which we call the `source', the island was formed from the film with $\Delta_1$ and the leads had the smaller gap, a configuration that reduces QP trapping on its island. Measurements of the switching current in this device as a function of gate charge, $Q_\textrm{src}$, have shown clean $2e$ periodicity, as expected in the absence of QP trapping. This source device is voltage-biased ($\sim$1 k$\Omega$ load-line impedance), and is shunted on chip by the $\sim$2 pF stray capacitance of its leads.

In the device on the left of Fig.~\ref{circuit}(a) (the `detector') the film with the lower gap $\Delta_2$ formed the island, and the leads had the higher gap $\Delta_1$. In this configuration the SCPT island can trap QPs far more effectively. We used the QP trapping and untrapping rates, which are very sensitive to electromagnetic noise in the environment and can be photo activated, Fig.~\ref{circuit}(b), to detect the radiation emitted by the source.

We configured the detector for rf reflectometry measurement of its charge-dependent Josephson inductance at zero dc bias \cite{Sillanpaa04,NaamanQP06}, and recorded the temporal variation of the reflected power, P$_\textrm{ref}$, which indicates the presence or absence of a single extra QP on the transistor's island, as shown in Fig.~\ref{circuit}(c) (details of the measurement appear in Ref.~\cite{NaamanQP06}). The signal was sampled at 100 ns intervals, and the typical time record spans 300 ms. The signal shows telegraph switching between two levels, with the upper level corresponding to the presence of an extra QP on the detector's island (odd parity state), and the lower one corresponding to its absence (even parity). We operate on the measured telegraph signal with a cumulative likelihood-ratio algorithm \cite{Lu03,Hinkley71} to discriminate between the two signal levels and find the dwell times of the system in the two states. 

Figure \ref{circuit}(d) shows the detector's even state probability $P_\textrm{even}$, as we biased the drain-source and gate voltages on the source device. For each point $\left\{Q_\textrm{src},V_\textrm{src}\right\}$ in the figure we recorded the statistics of QP tunneling in the detector and determined $P_\textrm{even}$ from the fraction of time spent with no extra quasiparticles on its island. It is immediately clear from the figure that $P_\textrm{even}$ changes significantly in response to the source's operating point, and exhibits a surprisingly intricate structure. $P_\textrm{even}$ changes with $V_\textrm{src}$ both along horizontal lines, independently of $Q_\textrm{src}$, and along sloped lines that are $Q_\textrm{src}$-dependent. In addition, $P_\textrm{even}$ is observed to \textit{increase} above its $V_\textrm{src}=0$ value ($\sim$\:12\:\%) [bright ridges, arrows in Fig.~\ref{circuit}(d)], a somewhat counter-intuitive result given that QP poisoning is generally assumed to be enhanced in the presence of electromagnetic noise \cite{Mannik04}. We observed similar behavior in all four pairs of devices that we measured. 

\begin{figure}[t]
\epsfxsize=3.2in
\epsfbox{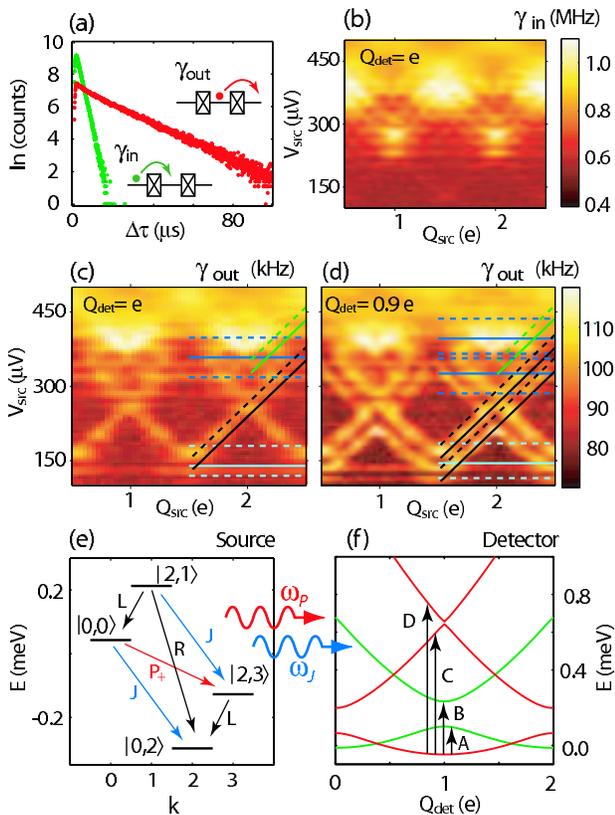}
\caption{\label{Q2V2} (color). (a) Lifetime histograms in the even (green) and odd (red) states. (b) $\gamma_\textrm{in}$ and (c) $\gamma_\textrm{out}$ vs.~source operating point at $Q_\textrm{det}=e$. (d) $\gamma_\textrm{out}$ at $Q_\textrm{det}=0.9\,e$. (e) Radiative cascades in the source emit microwave radiation that excites interband transitions (f) in the detector. Solid lines in (c),(d) correspond to $\hbar\omega_J=\delta E_{A,B}$ (light blue), $\hbar\omega_J=\delta E_{C,D}$ (dark blue), $\hbar\omega_{P_{-}}=\delta E_{C,D}$, $n=3$ (black), $\hbar\omega_{P_{-}}=\delta E_{B}$, $n=2$ (green). Dashed lines: mixing products with the detector's self resonance (see text).}
\end{figure}

Since $P_\textrm{even}$ depends on both in and out QP tunneling rates, $\gamma_\textrm{in}$ and $\gamma_\textrm{out}$, it does not provide sufficient information to understand the source\textendash detector interactions. However, having access to the QP tunneling dynamics in the time domain allows us to separate $P_\textrm{even}=\gamma_\textrm{out}/(\gamma_\textrm{out}+\gamma_\textrm{in})$ into the constituent rates.  Since we know the dwell times of the system in the even and odd states, we can histogram them to reveal the statistics of QP tunneling events into and out of the island, Fig.~\ref{Q2V2}(a). Using a procedure outlined in Ref.~\cite{Naaman06}, we determined the receiver response times $\tau_r^\textrm{out}=0.73\;\mu$s and $\tau_r^\textrm{in}=0.62\;\mu$s for the different transitions; using these numbers to account for the finite measurement bandwidth, we extract the actual tunneling rates from the observed lifetime histograms \cite{Naaman06,NaamanQP06}. The resulting tunneling rates are shown in Fig.~\ref{Q2V2} (b) and (c). Comparing Figs.~\ref{Q2V2} (b,c) and \ref{circuit} (d), we see that the observed increase in $P_\textrm{even}$ [arrows in Fig.~\ref{circuit} (d)] is due to enhancement of the QP untrapping rate $\gamma_\textrm{out}$ rather than a suppression of $\gamma_\textrm{in}$. 

We first discuss the enhancement of the QP escape rate $\gamma_\textrm{out}$ that appears to dominate the detector's response below $V_\textrm{src}$$\sim$200~$\mu$V. Figure \ref{Q2V2}(f) shows the energy band diagram for the detector device. The even parity bands (green) correspond to the ground and first excited states of the transistor, and the odd parity bands (red), are offset by the energy $\Delta_1-\Delta_2$ that is gained by a QP tunneling from the leads to the island \cite{Aumentado04}. We identify four possible transitions that, if excited, lead to enhanced QP escape rates. The processes labeled A and B in the figure directly transfer the extra QP from the island to the leads, leaving the SCPT in its ground and first excited state, respectively. Processes C and D do not directly change the parity of the SCPT; rather they excite the transistor to higher bands, from which the odd QP can escape spontaneously. The energy differences $\delta E_{A-D}$ for transitions A\textendash D depend on the detector charge, so that at different $Q_\textrm{det}$ the detector is sensitive to \emph{different sets} of frequencies. 

What emission processes in the source SCPT drive these transitions? It is convenient to represent the voltage-biased source by the states $|n,k\rangle$, where $n$ is the number of charges on the island, $k$ is the number of charges that have passed through the device, and the energy of the state $|n,k\rangle$ is given by $E_{n,k}=E_C(n-Q_\textrm{src}/e)^2-keV_\textrm{src}$ \cite{JoyezThesis,Siewert96}, as shown in Fig.~\ref{Q2V2}(e). As Cooper-pairs are transported through the device under voltage bias, the system cascades down the energy ladder, emitting microwave radiation into the environment; the linewidth of this radiation is set by voltage fluctuations across the device. The processes labeled J in the figure correspond to a transfer of a Cooper-pair (CP) across the whole device, producing the usual Josephson radiation $\hbar\omega_J=2eV_\textrm{src}$ independently of $Q_\textrm{src}$. Processes L,R transfer a single CP through a single junction, and process P$_+$ (P$_-$) transfers a CP across both junctions, simultaneously with the tunneling of an additional CP onto (off of) the island, emitting radiation at frequencies $\hbar\omega_{P_\pm}=3eV_\textrm{src}-4E_C[1\pm(n-Q_\textrm{src}/e)]$ . This radiative cascade in the source is analogous to that known from atomic physics \cite{JoyezThesis,CohenTannoudji}. 

Our model predicts that the source SCPT produces radiation everywhere in the $\{Q_\textrm{src},V_\textrm{src}\}$ space; its spectrum depends on the electrometer operating point, and contains the above frequencies. Our QP detector, however, will show enhanced QP tunneling rates only at those $\{Q_\textrm{src},V_\textrm{src}\}$ points where the source's emission frequencies are resonant with the detector's QP transitions. The loci of these points, estimated from the parameters of the two devices, are overplotted as solid lines in Fig.~\ref{Q2V2}(c), and in Fig.~\ref{Q2V2}(d), which shows $\gamma_\textrm{out}$ at $Q_\textrm{det}=0.9\:e$. Observe how transitions C and D, which are nearly degenerate at $Q_\textrm{det}=e$, become distinct at $Q_\textrm{det}=0.9\:e$, which results in the splitting of the `X'-shaped structures in Fig.~\ref{Q2V2}(d). We do not observe transitions that are due to the L and R processes in the source. Note that the $I-V$ characteristics of the detector (not shown) suggest that the device has a self-resonance with $\hbar\omega_r\sim80\;\mu e$V, likely associated with the geometry of its leads \cite{Holst94}; the dashed lines in Fig.~\ref{Q2V2}(c,d) correspond to the mixing products of this mode with the incoming radiation. 

The magnitude of the ac voltage induced on the detector by the radiation field, $V_{ac}$, that is required to produce the observed enhancement $\tilde{\gamma}_\textrm{out}$ of QP escape rates via processes C and D, can be estimated from perturbation theory \cite{CohenTannoudji}: $\tilde{\gamma}_\textrm{out}=(eV_{ac})^2(E_J/\hbar\omega_{C,D})^2/16\hbar^2\Gamma_0$, where $\Gamma_0=g_t\delta/4\pi\hbar$ is the spontaneous QP escape rate, $g_t$ is the dimensionless tunnel conductance and $\delta$ is the island level spacing \cite{Lutchyn06}. With $\omega_{C,D}\sim160$ GHz and the measured escape rate enhancement, we estimate $V_{ac}$ to be on the order of 50 nV. 

The activation of the rate $\gamma_\textrm{in}$, Fig.~\ref{Q2V2}(b), can be understood along similar lines. Since the odd state of the transistor is energetically favorable, QP trapping is spontaneous and $\gamma_\textrm{in}$ is limited by the QP density in the detector \cite{Aumentado04}. Radiation with frequencies $\hbar\omega\geq2\Delta_1,\:2\Delta_2$ can break CPs and increase the QP density, leading to a faster `poisoning' rate. Direct QP transitions are also possible if, for example, $\hbar\omega\geq\Delta_1+\Delta_2-\delta E^{eo}$, where $\delta E^{eo}$ is the difference in electrostatic energy between the even and odd states.  At voltages higher than 400 $\mu$V, the QP current in the source becomes significant through transport mechanisms that include Josephson-quasiparticle cycles \cite{Siewert96}, 3e processes \cite{Hadley98}, and sequential QP tunneling. This current may contribute to broadband  noise that appears to globally enhance both tunneling rates in the detector above $V_\textrm{src}$$\sim$\:400 $\mu$V. The individual rates become unmeasurably fast when the source is biased above its superconducting gap edge; this was also observed in Ref.~\cite{Mannik04}. 

While our model for the source--detector interactions does not explain all the features seen in Fig.~\ref{Q2V2}(b-d), it does account for the positions of the most prominent ones with good agreement and with no adjustable parameters. We emphasize that although our experiment is sensitive only to radiation at the relatively high frequencies of QP transitions, much lower frequencies approaching those of a typical qubit level splitting can be produced at electrometer bias points close to resonant CP tunneling lines. The intensity of the emitted radiation can be calculated, in principle, by solving the master equation for the biased transistor \cite{JoyezThesis}.

\begin{figure}[t]
\epsfxsize=3.2in
\epsfbox{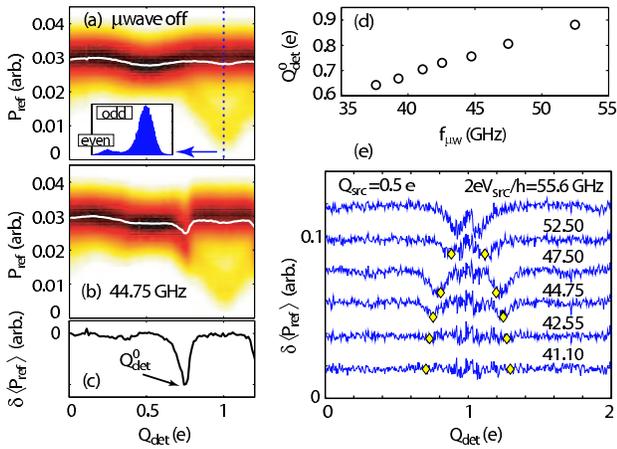}
\caption{\label{spect} (color online). (a),(b) Histograms of the telegraphic reflected power vs.~$Q_\textrm{det}$, (b) with and (a) without  microwave irradiation; white lines: the time-averaged reflected power $\langle$P$_\textrm{ref}\rangle$. The inset in (a) shows the histogram at $Q_\textrm{det}=e$. (c) Difference in the averaged reflection, $\delta\langle$P$_\textrm{ref}\rangle$ (d) Location of the dip in $\delta\langle$P$_\textrm{ref}\rangle$ vs.~microwave frequency $f_{\mu w}$. (e) $\delta\langle$P$_\textrm{ref}\rangle$ as a function of $Q_\textrm{det}$, for source voltages $2eV_\textrm{src}=hf_{\mu w}$ at $Q_\textrm{src}=0.5e$ (data offset for clarity). Diamonds: $Q^0_\textrm{det}(f_{\mu w})$ data of panel (d).}
\end{figure}

We further tested our interpretation of the data by studying the response of the detector to radiation with known frequency, applied by an external microwave generator. We spectroscopically mapped the transition frequencies of the detector by finding $Q_\textrm{det}$ at which the QP escape rate was enhanced for a given frequency $f_{\mu w}$ of the generator. The inset in Fig.~\ref{spect}(a) shows a histogram of the reflected probe signal over a 20 ms window at $Q_\textrm{det}=1$; the two peaks correspond to the two telegraph levels of the signal in the even and odd states. The intensity plot in the main panel of Fig.~\ref{spect}(a) represent a stack of these histograms, acquired at different $Q_\textrm{det}$ values in the absence of radiation. A similar plot is shown in Fig.~\ref{spect}(b) but with external radiation applied at $f_{\mu w}=44.75$ GHz. At this frequency the weight in the histogram shifts from the odd to the even peak at a particular $Q^0_\textrm{det}=0.76\:e$. This shift in weight [dip in Fig.~\ref{spect}(c)], essentially an increase in $P_\textrm{even}$, occurs at different values of $Q^0_\textrm{det}$ for different frequencies $f_{\mu w}$, as shown in Fig.~\ref{spect}(d). These values are symmetric about odd-integer gate charges and are $2e$ periodic. We found that the QP escape rate $\gamma_\textrm{out}$ at $Q^0_\textrm{det}$ grows linearly with microwave power (not shown), as expected from perturbation theory for photon-assisted tunneling. 

If we now generate radiation by voltage-biasing the source SCPT in place of the external generator and repeat the above measurements, the enhancement of the detector even-state probability should follow the same $Q^0_\textrm{det}(f_{\mu w})$ dependence of Fig.~\ref{spect}(d), where now the microwave frequency is given by $hf_{\mu w}=2eV_\textrm{src}$ (Josephson radiation). The results of this experiment are shown in Fig.~\ref{spect}(e), where we have plotted the time-averaged reflected probe signal, relative to the `dark' response, for a set of source voltages chosen to coincide with the frequencies in \ref{spect}(d). We see excellent agreement in $Q^0_\textrm{det}(f_{\mu w})$\textemdash the detector response to microwave radiation, whether its origin is Josephson radiation from the source device or external monochromatic microwave signal.

To conclude, we have detected narrow-band microwave radiation emitted by a voltage-biased SCPT electrometer by use of a nearby QP tunneling detector. We have identified the QP transition frequencies in the detector, and the emission processes in the electrometer\textemdash these include the usual Josephson radiation, as well as radiative cascade processes.  The radiation emitted by the electrometer, when coupled to other devices on the chip, may not only assist QP transitions as observed here, but also excite charge traps, defects, or higher energy levels in the device, effectively interfering with its proper operation. Therefore, we argue, care should be taken when using a biased SCPT for qubit readout in choosing the electrometer operating point and in engineering the high frequency coupling between the devices.   

We thank K.~W.~Lehnert for valuable discussions.


\begin{thebibliography}{28}
\bibitem[{Not()}]{NoteCoupling}
\bibinfo{note}{The power coupling between the two devices, likely frequency
  dependent, is not well characterized}.
\end{thebibliography}

\end{document}